\documentclass[a4paper,showpacs,prb,twocolumn]{revtex4}
\usepackage{amsfonts}
\usepackage{graphicx}
\usepackage{color}
\usepackage{amsmath}
\usepackage{amssymb}
\usepackage{latexsym}
\usepackage{psfrag}

\begin{document}

\title{Magnetic barriers in graphene nanoribbons: Theoretical study of transport properties}
\author{Hengyi Xu}
\author{T. Heinzel}
\email{thomas.heinzel@uni-duesseldorf.de}
\affiliation{Condensed Matter Physics Laboratory, Heinrich-Heine-Universit\"at,
Universit\"atsstr.1, 40225 D\"usseldorf, Germany}
\author{M. Evaldsson}
\author{I. V. Zozoulenko}
\email{Igor.Zozoulenko@itn.liu.se}
\affiliation{Solid State Electronics, Department of Science and Technology, Link\"{o}ping
University, 60174 Norrk\"{o}ping, Sweden}
\date{\today }

\begin{abstract}
A theoretical study of the transport properties of zigzag and armchair graphene
nanoribbons with a magnetic barrier on top is presented. The magnetic barrier modifies
the energy spectrum of the nanoribbons locally, which results in an energy shift of the
conductance steps towards higher energies. The magnetic barrier also induces
$\mathrm{{Fabry-P\acute{e}rot}}$ type oscillations, provided the edges of the barrier are
sufficiently sharp.  The lowest propagating state present in zigzag and metallic armchair
nanoribbons prevent confinement of the charge carriers by the magnetic barrier.
Disordered edges in nanoribbons tend to localize the lowest propagating state, which get
delocalized in the magnetic barrier region. Thus, in sharp contrast to the case of
two-dimensional graphene, the charge carriers in graphene nanoribbons cannot be confined
by magnetic barriers. We also present a novel method based on the Green's function
technique for the calculation of the magnetosubband structure, Bloch states and
magnetoconductance of the graphene nanoribbons in a perpendicular magnetic field.
Utilization of this method greatly facilitates the conductance calculations, because, in
contrast to excising methods, the present method does not require self-consistent
calculations for the surface Green's function.
\end{abstract}

\maketitle

\section{Introduction}

The single planar sheet with carbon atoms densely packed in a honeycomb
structure forms the so-called graphene, which demonstrates a variety of
unique electronic transport properties and has the potential applications in
the future nanoelectronics \cite{Neto}. Theoretical studies have indicated
that the special lattice structure of the graphene results in nearly linear
dispersion relations around the K points (Dirac points) of the Brillouin
zone \cite{Wakabayashi1999}. This unique band structure is responsible for
the distinct electronic properties of the graphene. Near the Dirac point,
electrons manifest themselves the massless chiral fermions and can be
described by the Dirac equation \cite{Katsnelson, Pereira, Tworzydlo}. The
electronic transport behaviors of the two-dimensional graphene subjected to
an electrostatic potential \cite{Katsnelson} or a magnetic barrier (MB) \cite{Martino} were studied on the basis of the Dirac equation, which indicate
that the Dirac fermions can be transmitted perfectly through a classically
forbidden region while confined effectively by the magnetic barrier.
Moreover, the anomalous integer and fractional quantum Hall effects in
two-dimensional graphene have been studied experimentally and theoretically
by various groups \cite{Zhang,Novoselov,Peres2006,Gusynin}.

The rolled-up graphene is known as the single-wall carbon nanotube whose electronic
properties have been studied extensively in the past decades. The quantized conductance
and $\mathrm{{Fabry-P\acute{e}rot}}$ interference pattern were observed experimentally and interpreted by
various theoretical approaches
\cite{Liang}. The other interesting effects including Coulomb blockade \cite{Herrero} and Kondo effects \cite{Nygard}, and the electronic transport in ballistic
\cite{Javey} and disordered nanotubes \cite{Hjort} were studied. Another related
carbon-based structure is the graphene nanoribbon (GNR), referred to the quasi-one
dimensional graphene with a finite width $W$. Recent development of the experimental
technique enable one to fabricate very narrow GNRs with ultrasmooth edges of the width
$W\leq 10$ nm\cite{Li}. The electrons propagate in such narrow systems very differently
compared with the two-dimensional graphene where the edges are totally irrelevant. In
graphene ribbons, the transport properties are strongly influenced by their edges along
the transport direction which are distinguished into two types: zigzag and armchair. For
armchair case, it is particularly interesting that the graphene ribbons may be metallic
or semiconducting depending on their widths. There is a lot of theoretical effort devoted
to the studies of the quantum transport in graphene ribbons. The conductance quantization
in mesoscopic graphene \cite{Peres2006} and coherent transport in graphene
nanoconstrictions with or without defects \cite{Munoz-Rojas} were reported recently.

The purpose of the present paper is twofold. First, we explore a possibility
to control electron conductance of graphene nanoribbons with the help of
magnetic barriers. MBs in the conventional quantum wires (QWRs) have been
the subject of theoretical and experimental studies, which are driven by the
MB's potential ability of parametric spin filtering. The pioneering
theoretical research by Peeters et al. \cite{Peeters} indicated that the
magnetic barrier possesses the wave-vector filtering properties in QWRs and
further work in graphene was also suggested \cite{Pereira}. Furthermore,
recent theoretical studies have revealed further rich
phenomenology of  magnetic barriers in quantum wires, such as Fano-type
resonances \cite{Xu} and spin filtering \cite{Majumdar1996,Zhai2006}. In two-dimensional
graphene, theoretical work has shown the strong effects of the magnetic
barrier on the direction-dependent transmission \cite{Martino}. Our studies
will focus on the magnetic barrier effects on the quasi-one-dimensional GNRs.

Second, we present a detailed description of a novel method based on the
Green's function technique for the calculation of the magnetosubband
structure, Bloch states and magnetoconductance of the graphene nanoribbons
in a perpendicular magnetic field. Note that magnetoconductance calculations
for the graphene nanoribbons based on the Green's function technique has
been reported previously \cite{Munoz-Rojas,Datta_Klimeck}. However, a
distinct feature of the present method is a novel approach to calculation of
the surface Green's function $\Gamma$ for semi-infinite nanoribbons. In
contrast to the Green's functions for finite structures that can be easily
calculated by adding slice by slice in a recursive way with the help of the
Dyson's equation, the calculation of the surface Green's function of a
semi-infinite structure represents a non-trivial problem. Such calculations
are typically done self-consistently which makes conductance calculations
very time-consuming. In the present paper we present a different method of
computing $\Gamma$ which \textit{does not require self-consistent
calculations}. Instead, the surface Green's function is expressed via the
Bloch states of the graphene nanoribbons which in turn are simply obtained
as solutions of the eigenequation of the dimension $2N\times 2N$ (with $N$
being the width of the nanoribbon). Utilization of this method greatly
facilitates the conductance calculations, making the present method far more
efficient in comparison to the existing ones. Programming codes for calculation of the magnetosubband structure,
the surface Greens function and the magnetoconductance based on the developed method are freely available in the AIP EPAPS electronic depository.\cite{EPAPS}\newline

This paper is organized as follows: In Sec. II we sketch the geometry of the
devices and briefly introduce the model for the conductance for our
calculations. In Sec. III, we describe the tight-binding model for the
graphene, theory of the Green's function method, as well as the formalism
for the computation of surface Green's functions. This is followed by the
presentation and discussion of the numerical results in Sec. IV. Summary and
conclusion constitute the Sec. V.

\section{Formulation of the problem}

The geometries under consideration for graphene nanoribbons with zigzag and
armchair edges are illustrated in Fig. 1(b) and (c), respectively, where the
left and right leads are made of semi-infinite graphene. The nanoribbons are
subjected to a magnetic barrier whose shapes may be rectangular or smooth as
shown in Fig. 1(a) with zero magnetic field in leads. (Note however that the
theory presented in the next section is not restricted to the case of zero
field in the leads). The magnetic barrier represents a strongly localized
magnetic field that is oriented perpendicular to the surface of the ribbon.
Magnetic barriers with amplitudes up to $1\,\mathrm{T}$ have been realized experimentally by ferromagnetic films on top of a graphene
sheet \cite{Kubrak2000,Cerchez2007}: magnetizing the ferromagnetic film in the transport direction results
in a magnetic fringe field with a perpendicular component localized at the
edge of the film that extends along the transverse direction. Alternatively, magnetic barrier formation has been demonstrated by placing two-dimensional electron gases with a step in an external magnetic field,\cite{Leadbeater1995} an approach which conceptually allows much larger barrier amplitudes. Both concepts should be in principle adaptable to graphene nanoribbons.

\begin{figure}[tbp]
\includegraphics[scale=1]{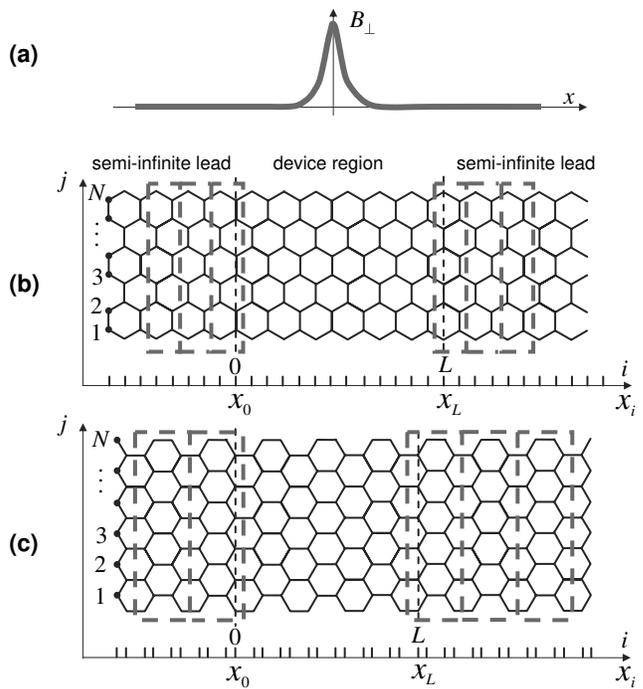} 
\caption{Schematic geometry of the structure under consideration for the
case of (b) zigzag and (c) armchair graphene. The current through the
central part of the device is injected and collected in semi-infinite ideal
leads representing graphene nanoribbons of width $N$. Unit cells of the
graphene nanoribbons are marked by blue dashed rectangles (see Fig. \protect
\ref{fig2:Bloch}). (a) A ferromagnetic film deposited on the top of the
graphene nanoribbons gives rise to an inhomogeneous magnetic field.}
\label{fig1:geometry}
\end{figure}

We model the leads and the device in the middle by the standard
tight-binding Hamiltonian on the honeycomb lattice, see below, Eq. (\ref%
{Hamiltonian}). The conductance $\mathbf{G}$ can be calculated using the
Landauer-B\"{u}ttiker formalism which gives the conductance of the system in
terms of the electron transmission coefficient $\mathcal{T}$, expressed as
\begin{equation}
\mathbf{G}=-\frac{2e^{2}}{h}\int dE\,\mathcal{T}(E)\frac{\partial
f_{FD}\left( E-E_{F}\right) }{\partial E},  \label{conductance}
\end{equation}%
where $\mathcal{T}(E)$ is the total transmission coefficient, $%
f_{FD}(E-E_{F})$ is the Fermi-Dirac distribution function and $E_{F}$ is the
Fermi energy.

We calculate the transmission amplitudes of electrons injected to the
systems using the recursive Green's function method which is described in
the next section.

\section{Theory}

\subsection{Basics}

\begin{figure}[tbp]
\includegraphics[scale=1]{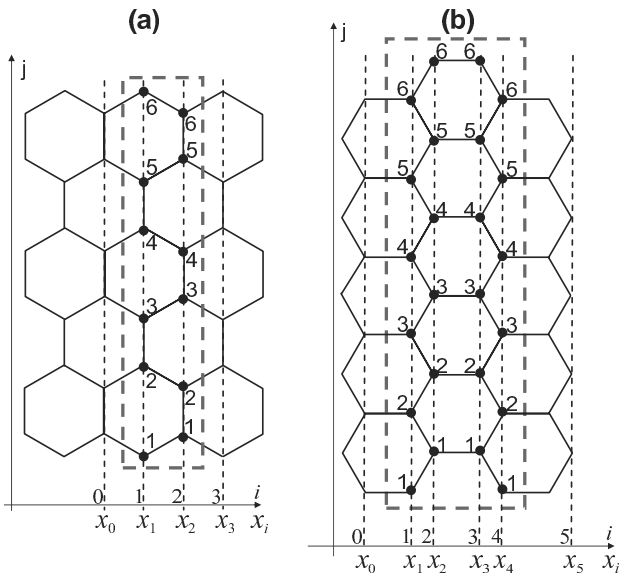} 
\caption{Geometry of (a) zigzag and (b) armchair graphene ribbons. The
ribbons are periodic in the $x$-direction (slices in the $x$-directions are
labeled by index $i$). The figure shows nanoribbons with $N=6$ sites in the
transverse direction. Unit cells of the graphene nanoribbons are marked by
dashed rectangles.}
\label{fig2:Bloch}
\end{figure}

We define the Bloch states in the infinite periodic graphene ribbons,
\begin{equation}
|\psi \rangle =\sum_{i,j}\psi _{i,j}a_{i,j}^{+}|0\rangle ,\;\psi
_{i,j}=e^{ikx_{i}}\varphi _{i,j},  \label{Bloch}
\end{equation}%
where $a_{i,j}^{+}\,(a_{i,j})$ is a standard creation (annihilation)
operator on the site $(i,j)$; $\psi _{i,j}$ is the amplitude of the wave
function on the site $(i,j);$ $x_{i}$ is the coordinate of the $i$-th slice,
$k$ is the Bloch wave vector in the direction of the translational
invariance $x$, and the summation runs over all sites of the graphene
lattice (see Fig. \ref{fig2:Bloch}). Note that this form of the wave
function does not distinguish between sublattices A and B of the graphene
lattice. An explicit distinction between these sublattices is not necessary
when using the Green's function technique, where, instead, it is more
convenient to define the wave function on slices of the lattice (see Sec. %
\ref{Sec:Bloch states}).

The standard tight-binding Hamiltonian has the form%
\begin{equation}
H=\sum_{r}V_{r}a_{r}^{+}a_{r}-\sum_{r,\Delta }t_{r,r+\Delta
}a_{r}^{+}a_{r+\Delta },  \label{Hamiltonian}
\end{equation}%
where $V_{r}$ describes the electrostatic potential on the site $r=i,j$ and
summation in the second term is performed over all available nearest
neighbors with $t_{r,r+\Delta }$ being the nearest-neighbor hopping
integral. In the absence of a magnetic field the nearest-neighbor hopping
integral is $t_{r,r+\Delta }=t_{0}\approx 2.7$ eV. In the presence of an
external perpendicular magnetic field $B$ the hopping integral acquires the
Peierls phase factor, $t_{r,r+\Delta }=t_{0}\exp (i\theta _{r,r+\Delta }),$%
where $\theta _{r,r+\Delta }=2\pi \phi _{r,r+\Delta }/\phi _{0}$ , with $%
\phi _{r,r+\Delta }$ being the line integral of the vector potential $%
\mathbf{A}$ from site $r$ to a neighboring site $r+\Delta ,$
\begin{equation}
\phi _{r,r+\Delta }=\int_{r}^{r+\Delta }\mathbf{A}\cdot d\mathbf{l}
\label{Phase}
\end{equation}%
and $\phi _{0}=h/e$ is the flux quantum (in our calculations we use the
Landau gauge, $\mathbf{A}=(-By,0)).$ (In calculation of hopping integral (%
\ref{Phase}) we use the carbon-carbon bond length $a=0.142$ nm, see
Appendix). Note that the Hamiltonian operator $H$ is convenient to write
down in the form,%
\begin{equation}
H=\sum_{i}\left[ h_{i}\right] +U,  \label{H_U}
\end{equation}%
where $h_{i}$ describes the Hamiltonian of the $i$-th slice, and $U$
describes hopping between all neighboring slices (explicit forms of $h_{i}$
and $U$ can be easily obtained from Eq. (\ref{Hamiltonian})).

The Green's function of the operator $H$ is defined in a standard way\cite%
{Datta_book,Ferry},
\begin{equation}
\left( E-H+i\varepsilon \mathcal{\,}\right) G=I  \label{Green}
\end{equation}%
where $I\mathcal{\,}$ is the unitary operator.

\subsection{Bloch states and velocities in the graphene nanoribbons.}

\label{Sec:Bloch states}

We continue by describing a method for calculation of the Bloch states and
their group velocities in the zigzag and armchair graphene nanoribbons in
the presence of a perpendicular magnetic field. The method is based on the
technique developed for calculation of the band structure of a mesoscopic
antidot lattice in confined geometries \cite{Z} and has been used for
calculation of the Bloch states in photonic structures \cite{Rahachou} and
in the interacting quantum wires in the integer quantum Hall regime \cite%
{wires}.

Consider an infinite ideal graphene ribbon with $N$ sites in the transverse $%
j$-direction, Fig. \ref{fig2:Bloch}. A unit cell of the structure consists
of $M$ slices, where $M=2$ for the zigzag graphene and $M=4$ for the
armchair graphene.

The Hamiltonian of an ideal infinitely long graphene ribbon can be written
in the form
\begin{equation}
H=H_{\mathrm{cell}}+H_{\mathrm{out}}+U,\mathcal{\,}  \label{L_cell}
\end{equation}%
where the operators $H_{\mathrm{cell}}$ and $H_{\mathrm{out}}$ describe
respectively the unit under consideration ($1\leq i\leq M$), and the outside
region including all other slices $-\infty <i\leq 0$ and $M+1\leq i<\infty $%
, and $U\mathcal{\,}$ is the hopping operator between the cell and slices $%
i=0$ and $i=M+1$ (an explicit form for these operators can be easily
obtained from Eq. (\ref{Hamiltonian})). We write the total wave function,
Eq. (\ref{Bloch}), in the form%
\begin{equation}
|\psi \rangle =|\psi _{\mathrm{cell}}\rangle +|\psi _{\mathrm{out}}\rangle ,
\label{psi_cell}
\end{equation}%
where $|\psi _{\mathrm{cell}}\rangle $ and $|\psi _{\mathrm{out}}\rangle $
are respectively wave functions in the cell and in the outside region.
Substituting Eqs. (\ref{L_cell}),(\ref{psi_cell}) into the Schr\"{o}dinger
equation $H|\psi \rangle =E|\psi \rangle $ and using the definition of the
Green's function, Eq. (\ref{Green}), we obtain $|\psi _{\mathrm{cell}%
}\rangle =G_{\mathrm{cell}}U|\psi _{\mathrm{out}}\rangle ,$ where $G_{%
\mathrm{cell}}$ is the Green's function of the operator $H_{\mathrm{cell}}$.
Taking the matrix elements of the wave functions in the real space
representations, $\psi _{i,j}=\langle 0a_{i,j}|\psi \rangle $ for the first (%
$i=1$) and the last ($i=M$) slices of the unit cell, this equation can be
written in the matrix form, 
\begin{align}
\psi _{1}& =G_{\mathrm{cell}}^{1,1}U_{1,0}\psi _{0}+G_{\mathrm{cell}%
}^{1,M}U_{1,0}^{+}\psi _{M+1}  \label{psi_cell_2} \\
\psi _{M}& =G_{\mathrm{cell}}^{M,1}U_{1,0}\psi _{0}+G_{\mathrm{cell}%
}^{M,M}U_{1,0}^{+}\psi _{M+1},  \notag
\end{align}%
where $\psi _{i}$ is the vector column describing the wave function for the
slice $i$, 
\begin{equation}
\psi _{i}=\left( \psi _{i,1};\ldots ;\psi _{i,N}\right) ^{T}  \label{psi_i}
\end{equation}%
and $U_{1,0}$ and $G_{\mathrm{cell}}^{i,i^{\prime }}$ denote the matrixes
with the matrix elements
\begin{eqnarray}
\left( U_{1,0}\right) _{jj^{\prime }} &=&\langle 0a_{1,j}|U|a_{0,j^{\prime
}}^{+}0\rangle  \label{matrix elements} \\
\left( G_{\mathrm{cell}}^{i,i^{\prime }}\right) _{jj^{\prime }} &=&\langle
0a_{i,j}|G_{\mathrm{cell}}|a_{i^{\prime },j^{\prime }}^{+}0\rangle .  \notag
\end{eqnarray}%
Explicit expressions for the matrix elements of the matrix $U$ are given in
the Appendix. In the derivation of Eq. (\ref{psi_cell_2}) we used $%
U_{M,M+1}=U_{0,1}$ (because of the periodicity of the ribbons) and $%
U_{0,1}=U_{1,0}^{+}$ (`$+$' stands for Hermitian conjugate).

It is convenient to rewrite Eq. (\ref{psi_cell_2}) in a compact form
\begin{align}
& T_{1}\left(
\begin{array}{c}
\psi _{M+1} \\
\psi _{M}%
\end{array}%
\right) =T_{2}\left(
\begin{array}{c}
\psi _{1} \\
\psi _{0}%
\end{array}%
\right) ,\;\mathrm{where}  \label{T1_T2} \\
& T_{1}=%
\begin{pmatrix}
-G_{\mathrm{cell}}^{1,M}U_{1,0}^{+} & \;0 \\
-G_{\mathrm{cell}}^{M,M}U_{1,0}^{+} & \;I%
\end{pmatrix}%
,\;T_{2}=%
\begin{pmatrix}
-I & \;G_{\mathrm{cell}}^{1,1}U_{1,0} \\
0 & \;G_{\mathrm{cell}}^{M,1}U_{1,0}%
\end{pmatrix}
\notag
\end{align}%
\newline
with $I$ being the unitary matrix. The wave function of the periodic
structure has the Bloch form,
\begin{equation}
\psi _{m+M}=e^{ikM}I\psi _{m}.  \label{Bloch_2}
\end{equation}%
Combining Eqs. (\ref{T1_T2}) and (\ref{Bloch_2}), we arrive at the
eigenequation,%
\begin{equation}
T_{1}^{-1}T_{2}\left(
\begin{array}{c}
\psi _{1} \\
\psi _{0}%
\end{array}%
\right) =e^{ikM}\left(
\begin{array}{c}
\psi _{1} \\
\psi _{0}%
\end{array}%
\right)  \label{Bloch_eigenequation}
\end{equation}%
determining the set of Bloch eigenvectors $k_{\alpha }$ and eigenfunctions $%
\psi ^{\alpha },$ $1\leq \alpha \leq N.$ It should be stressed that this
eigenequation provides a set of the Bloch states $\left\{ k_{\alpha
}\right\} $ for a fixed energy $E$, which includes both propagating and
evanescent states. The latter can be easily identified by a non-zero
imaginary part.

In order to separate right- and left-propagating states, $k_{\alpha }^{+}$
and $k_{a}^{-},$ we compute the group velocities of the Bloch states $%
v_{\alpha }=\frac{\partial E}{\partial k_{\alpha }}$, whose signs determine
the direction of propagation (`+' stands for the right-propagating and `-'
for the left propagating states). The group velocities can be computed
directly by numerical differentiation of the dispersion relation. This is
however not an efficient approach because for each energy the eigensolver
gives eigenstates $\alpha $ in different order. We instead derive below a
simple formula which gives the group velocities of the Bloch states based on
the eigenfunctions of Eq. (\ref{Bloch_eigenequation}).

Consider a unit cell of an infinite graphene nanoribbon consisting of $M$
slices. The wavefunction of the $\alpha $-th Bloch state (\ref{Bloch}) can
be conveniently rewritten in the form $|\psi \rangle =\sum_{i=1}^{M}|\psi
_{i}\rangle ,$ where $|\psi _{i}\rangle $ is the wave function for the $i$%
-th slice,
\begin{equation}
|\psi _{i}\rangle =e^{ikx_{i}}|\varphi _{i}\rangle  \label{psi_phi}
\end{equation}%
[To simplify our notations we have dropped the Bloch index $\alpha $ ].
Starting from the Schr\"{o}dinger equation and calculating the matrix
element of the Hamiltonian of the unit cell, we obtain for each slice $i,$ $%
\langle \psi _{i}|H|\psi \rangle =E\langle \psi _{i}|\psi \rangle =E|\varphi
_{i}|^{2}.$ Performing summation over all slices of the unit cell and using
a definition of the group velocity, we obtain
\begin{equation}
v=\frac{\partial E}{\partial k}=\frac{1}{M}\sum_{i=1}^{M}\frac{\partial }{%
\partial k}\left[ \frac{\langle \psi _{i}|H|\psi \rangle }{|\varphi _{i}|^{2}%
}\right]
\end{equation}%
where the summation is performed over all slices $i$ of the unit cell, and
\begin{equation}
\varphi _{i}=(\varphi _{i,1};...;\varphi _{i,N})^{T}  \label{phi_vector}
\end{equation}%
is a vector composed of the matrix elements $\varphi _{i,j}=$ $\langle
0a_{i,j}|\varphi \rangle .$(Note that according to Eq. (\ref{psi_i}) and (%
\ref{psi_phi}), vectors $\varphi _{i}$ can be obtained from $\psi _{i}$ via
the relation $\psi _{i}=e^{ikx_{i}}\varphi _{i}$). Representing the
Hamiltonian of the unit cell in the form (\ref{H_U}), the matrix elements $%
\langle \psi _{i}|H|\psi \rangle $ can be easily evaluated, which gives%
\begin{eqnarray}
v &=&\frac{-i}{M}\sum_{i=1}^{M}\frac{\varphi _{i}^{\ast T}}{|\varphi
_{i}|^{2}}\Big[(x_{i}-x_{i-1})U_{i,i-1}\varphi _{i-1}e^{-ik(x_{i}-x_{i-1})}
\notag \\
&&-(x_{i+1}-x_{i})U_{i,i+1}\varphi _{i+1}e^{-ik(x_{i+1}-x_{i})}\Big],
\label{velocity}
\end{eqnarray}%
where the matrixes $U_{i,i^{\prime }}$ are defined by Eq. (\ref{matrix
elements}) [explicit expressions for these matrix elements are given in the
Appendix].

\subsection{Surface Greens function $\Gamma .$}

Here, we describe an efficient method for calculation of the surface Green's
function $\Gamma $ in the magnetic field.\cite{EPAPS} Note that most of the methods for
calculation of the Green's function reported to date require searching for a
self-consistent solution for $\Gamma $ which makes these calculation very
time consuming\cite{Munoz-Rojas,Datta_Klimeck}. In contrast, our method does
not require self-consistent calculations, and the surface Greens function is
simply given by multiplication of matrixes composed of the Bloch states of
the graphene lattice (see below, Eqs. (\ref{Gamma_r}),(\ref{Gamma_l})). The
calculations described in this section are based on the method developed in
Ref. \cite{Rahachou} for periodic photonic crystals which is adapted here
for the case of the graphene nanoribbons.

Consider a semi-infinite periodic ideal graphene ribbon extended to the
right in the region $-m\leq i<\infty $. Suppose that an excitation $%
|s\rangle $ is applied to its surface slice $i=-m$. Introducing the Green's
function of the semi-infinite ribbon, $G_{\mathrm{rib}},$ one can write down
the response to the excitation $|s\rangle $ in a standard form\cite%
{Datta_book}
\begin{equation}
|\psi \rangle =G_{\mathrm{rib}}|s\rangle ,  \label{psi}
\end{equation}%
where $|\psi \rangle $ is the wave function that has to satisfy the Bloch
condition (\ref{Bloch}). Consider a unit cell of a graphene lattice, $1\leq
i\leq M,$ ($M=2$ and $4$ for the zigzag and armchair lattices, see Fig. \ref%
{fig2:Bloch}). Applying Dyson's equation between the slices $0$ and $1$ we
obtain%
\begin{equation}
G_{\mathrm{rib}}^{1,-m}=\Gamma _{r}U_{1,0}G_{\mathrm{rib}}^{0,-m},
\label{G_psi}
\end{equation}%
where $\Gamma _{r}\equiv G_{\mathrm{rib}}^{1,1}$ is the right surface Green's function
(i.e. the surface function of the semiinfinite ribbon open to the right), and the
definition of the matrixes $U$ and $G$ in the real space representation are given by Eq.
(\ref{matrix elements})). Evaluating the matrix elements $\langle 0a_{1,j}|\psi \rangle $
of Eq. (\ref{psi}) and
making use of Eq. (\ref{G_psi}), we obtain for an each Bloch state $\alpha $%
, $\psi _{1}^{\alpha }=$ $\Gamma _{r}U_{1,0}\psi _{0}^{\alpha }.$ The latter
equations can be used for determination of $\Gamma _{r}$,%
\begin{equation}
\Gamma _{r}U_{1,0}=\Psi _{1}\Psi _{0}^{-1},  \label{Gamma_r}
\end{equation}%
where $\Psi _{1}$ and $\Psi _{0}$ are the square matrixes composed of the
matrix-columns $\psi _{1}^{\alpha }$ and $\psi _{0}^{\alpha },$ ($1\leq
\alpha \leq N),$ Eq. (\ref{Bloch_eigenequation}), i.e. $\Psi _{1}=(\psi
_{1}^{1},...,\psi _{1}^{N});$ $\Psi _{0}=(\psi _{0}^{1},...,\psi _{0}^{N}).$
The expression for the left surface Greens function $\Gamma _{l}$ (i.e. the
surface function of the semiinfinite ribbon open to the right) is derived in
a similar fashion,%
\begin{equation}
\Gamma _{l}U_{1,0}^{+}=\Psi _{M}\Psi _{M+1}^{-1},  \label{Gamma_l}
\end{equation}%
where the matrixes $\Psi _{M}$ and $\Psi _{M+1}$ are defined in a similar
way as $\Psi _{1}$ and $\Psi _{0}$ above. Note that matrixes $\Psi _{M}$ and
$\Psi _{M+1}$ can be easily obtained from $\Psi _{1}$ and $\Psi _{0}$ using
the relation (\ref{T1_T2}). Note also that when the magnetic field is
restricted to zero, the right and left surface Greens functions are
identical, $\Gamma _{l}=\Gamma _{r}.$

\subsection{Magnetoconductance of the graphene nanoribbons}

In order to calculate the transmission coefficient $\mathcal{T}(E)$ we
divide the structure into three regions, two ideal semi-infinite leads of
the width $N$ extending in the regions $i\leq 0$ and $i\geq L$ respectively,
and the central device region (where scattering occurs), see Fig. \ref%
{fig1:geometry}. We assume that the left and right leads are identical. The
incoming, transmitted and reflected states in the leads, $|\psi _{\alpha }^{%
\mathrm{i}}\rangle ,$ $|\psi _{\alpha }^{\mathrm{t}}\rangle $ and $|\psi
_{\alpha }^{\mathrm{r}}\rangle $, have the Bloch form (\ref{Bloch}),
\begin{align}
|\psi _{\alpha }^{\mathrm{i}}\rangle & =\sum_{i\leq 0}e^{ik_{\alpha
}^{+}x_{i}}\sum_{j=1}^{N}\phi _{i,j}^{\alpha }\,a_{i,j}^{+}|0\rangle
\label{i} \\
|\psi _{\alpha }^{\mathrm{t}}\rangle & =\sum_{i\geq L}\sum_{\beta }t_{\beta
\alpha }e^{ik_{\beta }^{+}(x_{i}-x_{L})}\sum_{j=1}^{N}\phi _{i,j}^{\beta
}\,a_{i,j}^{+}|0\rangle  \label{t} \\
|\psi _{\alpha }^{\mathrm{r}}\rangle & =\sum_{i\leq 0}\sum_{\beta }r_{\beta
\alpha }e^{ik_{\beta }^{-}x_{i}}\sum_{j=1}^{N}\phi _{i,j}^{\beta
}\,a_{i,j}^{+}|0\rangle ,  \label{r}
\end{align}%
where $t_{\beta \alpha }\,(r_{\beta \alpha })$ stands for the transmission
(reflection) amplitude from the incoming Bloch state $\alpha $ to the
transmitted (reflected) Bloch state $\beta ,$ and we choose $x_{0}=0.$ The
transmission and reflection coefficients are expressed through the
corresponding amplitudes and the Bloch velocities\cite{Datta_book}%
\begin{equation*}
\mathcal{T}=\sum_{\alpha ,\beta }\frac{v_{\beta }}{v_{\alpha }}|t_{\beta
\alpha }|^{2};\;\mathcal{R}=\sum_{\alpha ,\beta }\frac{v_{\beta }}{v_{\alpha
}}|r_{\beta \alpha }|^{2},
\end{equation*}%
where the summation runs over propagating states only. The transmission and
reflection amplitudes can be calculated from the equations\cite{Rahachou},
\begin{align}
\Phi _{1}T& =-G^{L,0}(U_{0,1}\Phi _{1}K_{1}-{\Gamma _{l}}^{-1}\Phi _{0})
\label{T} \\
\Phi _{0}R& =-G^{0,0}(U_{0,1}\Phi _{1}K_{1}-{\Gamma _{l}}^{-1}\Phi
_{0})-\Phi _{0}  \label{R}
\end{align}%
where the matrixes $T$ and $R$ of the dimension $N\times N_{prop}$ are
composed of the transmission and reflection amplitudes $(T)_{\beta \alpha
}=t_{\beta \alpha },$ $(R)_{\beta \alpha }=r_{\beta \alpha };$ (with $%
N_{prop}$ being the number of propagating modes in the leads); $G^{L,0}$ and
$G^{0,0}$ are the Green's function matrixes with matrix elements defined
according to Eq. (\ref{matrix elements}); $\Gamma _{l}$ is the left surface
Green's function, Eq. (\ref{Gamma_l}); $U_{0,1}$ is the hopping matrix
between the left lead and the device region (\ref{matrix elements}); $K_{1}$
is the diagonal matrix with the matrix elements $(K_{1})_{\alpha ,\beta
}=\exp (ik_{\alpha }^{+}\,x_{1})\delta _{\alpha ,\beta }.$ The square
matrixes $\Phi _{1}$ and $\Phi _{0}$ describe the Bloch states on the slices
1 and 0 of a ribbon unit cell (see Fig. \ref{fig2:Bloch}) and are composed
of matrix-columns $\phi _{1}^{\alpha }$ and $\phi _{0}^{\alpha },$ ($1\leq
\alpha \leq N),$ Eq. (\ref{phi_vector}), i.e. $\Phi _{1}=(\phi
_{1}^{1},...,\phi _{1}^{N});$ $\Phi _{0}=(\phi _{0}^{1},...,\phi _{0}^{N}).$

Calculation of the Green's functions $G^{L,0}$ and $G^{0,0}$ is performed in
a standard way \cite{Ferry}. We start from the Greens function of the first
slice in the device region and, using the Dyson's equation, add recursively
slice by slice until the last slice of this region is reached. Finally, we
apply the Dyson's equation two more times adding the left and right
semi-infinite ribbons whose surface Green's functions are given by Eqs. (\ref%
{Gamma_r}),(\ref{Gamma_l}).

Having calculated the transmission and reflection amplitudes that give the
wave functions on slices $i=0$ and $i=L,$ we can easily restore the wave
function inside the device region using the relation between the wave
functions on slices $i,i^{\prime }$ and $i+1,i^{\prime }-1$ (we assume that $%
i^{\prime }>i$) 
\begin{align}
\psi _{i+1}& =G_{\mathrm{inner}}^{i+1,i+1}U_{i+1,i}\psi _{i}+G_{\mathrm{inner%
}}^{i+1,i^{\prime }-1}U_{i^{\prime },i^{\prime }-1}^{+}\psi _{i^{\prime }}
\label{relation} \\
\psi _{i^{\prime }-1}& =G_{\mathrm{inner}}^{i^{\prime }-1,i+1}U_{i+1,i}\psi
_{i}+G_{\mathrm{inner}}^{i^{\prime }-1,i^{\prime }-1}U_{i^{\prime
},i^{\prime }-1}^{+}\psi _{i^{\prime }},  \notag
\end{align}%
where $G_{\mathrm{inner}}^{l,m}$ is the Green's function of the internal
region only (extending from the slice $i$ to the slice $m).$ (Equation (\ref%
{relation}) is derived in a similar way as Eq. (\ref{psi_cell_2})). Removing
slice by slice from the inner region and repeatedly using Eq. (\ref{relation}%
) on each step, we restore the wave function in the entire region $0<i<L.$

The diagonal elements of the total Green's function for each slice $i$ give
the local density of states (LDOS) at the site $i,j,$\cite{Datta_book} $\rho
(i,j,E)=-\frac{1}{\pi }\Im \left[ \left( G^{i,i}\right) _{jj}\right] .$ The
LDOS can be used to calculate the local electron density at the site $i,j$,
\begin{equation}
n(i,j)=\int dE\,\rho (i,j,E)f(E-E_{F}).
\end{equation}%
For quasi-one dimensional structures considered in this study it is
convenient to introduce the local density of states integrated in the
transverse direction,

\begin{equation}
\rho (i,E)=\sum_{j=1}\rho (i,j,E).  \label{LDOS}
\end{equation}

\section{Results and discussion}

\begin{figure*}[tbp]
\includegraphics[scale=1.2]{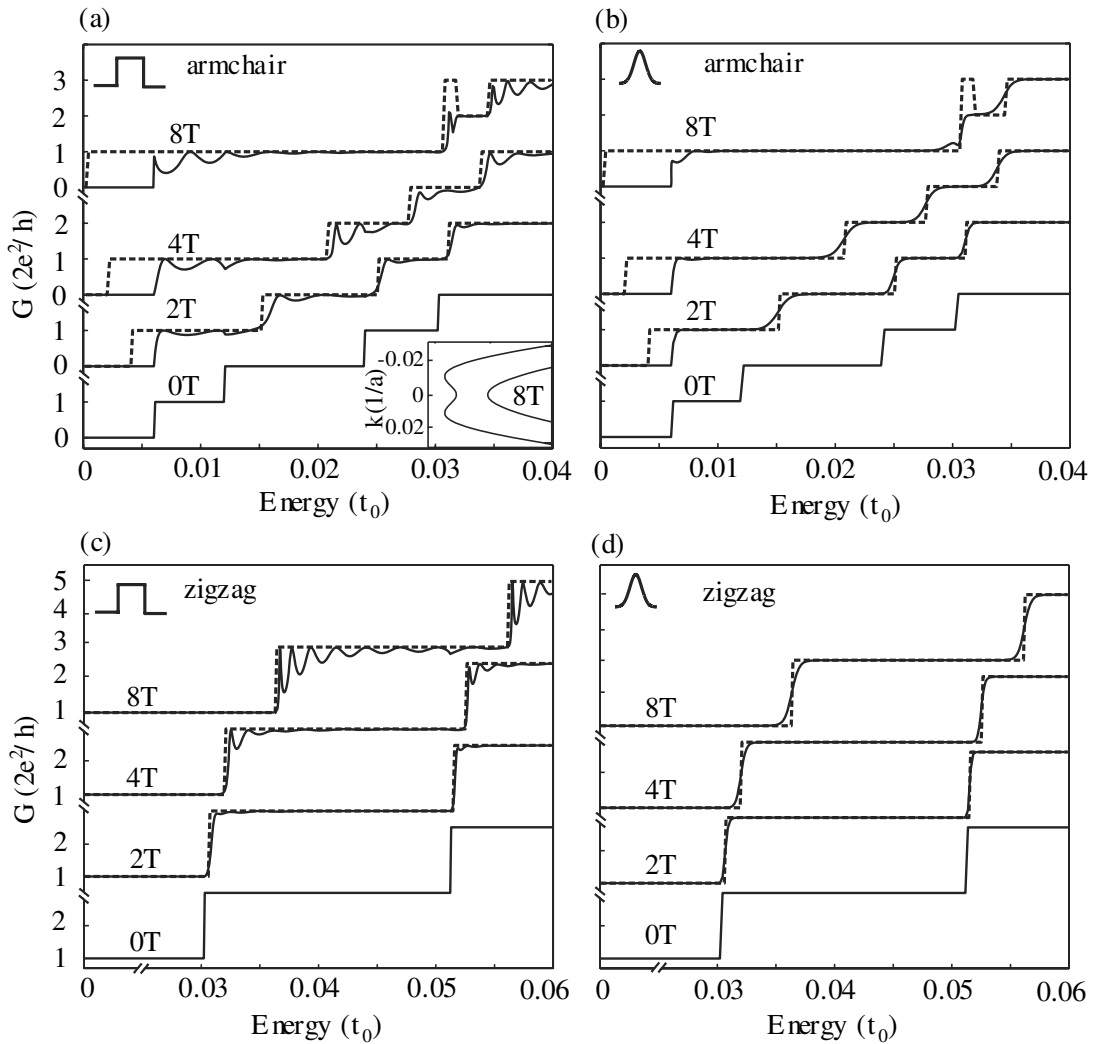} 
\caption{The calculated conductance (full lines) and the number of occupied modes at the
maximum magnetic field (dashed lines) as a function of the Fermi energy in the
semiconducting armchair GNR (a,b) and the zigzag GNR (c,d)
for MBs with different amplitudes $0\,\mathrm{T}$, $2\,\mathrm{T}$, $4\,\mathrm{T}$, and $8\,%
\mathrm{T}$. The inset in (a) represents the section of the energy
dispersion at $B=8\,\mathrm{T}$ which causes the change of the number of modes from 3 to
2 and back to 3 as the energy increases.} \label{fig3:Conductance}
\end{figure*}

In this section we discuss the conductance properties of two-terminal GNRs with MBs using
the formalism described above. GNRs with both zigzag and armchair edges are considered.
The electronic properties of armchair GNRs depend strongly on its width $W$. The armchair
GNRs are metallic when $2N+1$ is a multiple of 3, and otherwise they are semiconducting.
Metallic armchair GNRs behave similarly to zigzag GNRs regarding the effects discussed
here, even though the origin of the first subband is different, and are not presented
separately.

Figure \ref{fig3:Conductance} shows the Fermi energy
dependence of the conductance for the zigzag and armchair ribbons with $%
N=151 $ and $150$, respectively, corresponding to a width of $W\approx 32\,%
\mathrm{nm}$. The rectangular magnetic barrier has a length of $120\,\mathrm{%
nm}$. The smooth
magnetic barrier has the standard shape realized in experiments \cite
{Vancura2000,Cerchez2007} and a full width at half maximum of $120\,\mathrm{%
nm}$. For the case of
the smooth barrier the central (device) region has a length of $360\,\mathrm{%
nm}$. The shapes of the smooth and sharp barriers are depicted schematically in the
insets to Fig. \ref{fig3:Conductance}. We present the conductance
calculations for the maximum magnetic field strength in the barrier in the interval of
$0-8\,\mathrm{T}$. While inhomogeneous fields up to $\approx 3.4\,\mathrm{T}$ have been
achieved in the laboratory by using etch facets,\cite{Leadbeater1995} we consider such
high fields in order to address the regime when the magnetic length $l_B=\sqrt{\hbar/eB}$
($=26\,\mathrm{nm}$ at $1\,\mathrm{T}$) is smaller than the ribbon width. Alternatively,
this could have been achieved by increasing the ribbon width, which is however rather
impractical from the computational point of view.

In the absence of MBs, the ballistic conductance of the GNRs is simply proportional to
the number of subbands $N_{0}$ at the Fermi energy at zero magnetic field, \cite{Peres2006a} see Fig.
\ref{fig3:Conductance}. The conductance shows
plateaus and increases as a function of Fermi energy, in analogy to the case of QWRs.

Figures \ref{fig3:Conductance}(a), (b) show the conductance of the semiconducting
armchair GNR for the rectangular and smooth magnetic barriers. The dashed lines indicate
the number of propagating states $N_{B}$ in the corresponding GNR in the homogeneous
magnetic field whose amplitude is equal to the maximum field $B$ in the barrier region.
As the magnetic field increases the subbands depopulate and hence the corresponding
number of available propagating states $N_{B}$ decreases. Because the magnetic field
provides an additional confinement in the ribbon, at a given Fermi
energy the number of the magnetosubbands $N_{B}$ is always smaller than $%
N_{0}$. Because of this, $N_{B}$ represents the limiting factor for the
conductance of the magnetic barrier structure such that $N_{0}$ incoming
states in the leads are redistributed among $N_{B}$ available states in the
magnetic barrier. This is clearly seen in Figs. \ref{fig3:Conductance} (a),
(b) where the conductance of the structure at hand approximately follows $%
N_{B}.$ Note that the magnetic field reduces the energy gap in the vicinity of $E=0.$
Despite of this the conductance of the magnetic barrier is always zero below the energy
threshold of $E_{th}\approx 0.006t_{0}$ regardless of the strength of the magnetic
barrier.  This simply reflects the fact that propagating states are injected from the
leads where the magnetic field is absent and the threshold propagation energy
$E_{th}\approx 0.006t_{0}$ is not affected by the strength of the barrier in the central
region of the device.  In addition, transmission resonances are superimposed on the
conductance plateaus. They are well pronounced for the rectangular barriers, but get
heavily suppressed as the as magnetic barrier assumes the more realistic, soft shape. As
the strength of the barrier increases, the resonances become more prominent.

\begin{figure}[tbp]
\includegraphics[scale=1]{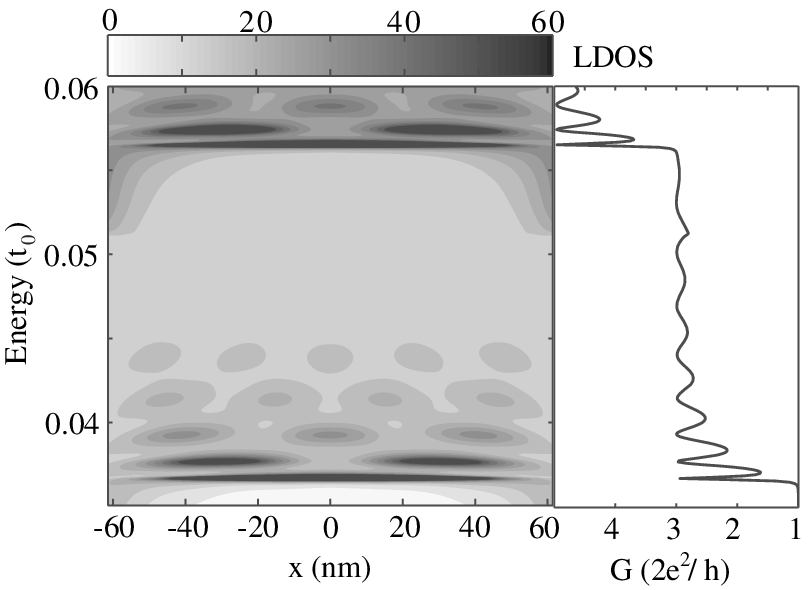}
\caption{The local density states (LDOS) with the rectangular MB of 8T integrated over
the y direction (in units of $t^{-1}_0$nm$^{-1}$), as a function of energy and $x$ in the
zigzag GNR. The LDOS plot shows the localized states form inside the magnetic barrier.
The position of the localized states in relation to the corresponding conductance.}
\label{fig4:LDOSxE}
\end{figure}

In the zigzag GNR with a MB, the conductance steps also move towards higher
energies and follow $N_{B}$ vs energy as $B$ increases, see Fig. \ref%
{fig3:Conductance}(c), (d).  This, as in the case of the armchair GNRs, simply reflects
the magnetic field induced shifts of the GNR modes in the barrier region. Around $E=0$ ,
an energy interval exists in which only the lowest propagating state contributes to the
conductance. This state evolves from the dispersionless edge state present in the zigzag
GNRs at zero energy. The MB is thus able to reduce the number of current carrying states
in certain energy intervals, e.g. between $E\approx 0.03t_{0}$ and $E\approx 0.037t_{0}$
for the MBs with a strengths of $8\,\mathrm{T}$. Note that for the zigzag
GNR the conductance changes in steps of $2\times2e^2/h$, whereas for the armchair GNR it
changes in steps of $2e^2/h$. This reflects the difference in evolution of the subband
structure of corresponding homogeneous armchair and zigzag GNRs, where the number of
states at the given energy depends on the wire width $N$ and on whether the ribbon is
metallic or insulating. (The conductance quantization for armchair and zigzag GNRs was
discussed by Peres \textit{et al.}).\cite{Peres2006a}

In addition, as in the case of the armchair GNR, transmission resonances are observed for
the rectangular MBs. These resonances are completely suppressed for the case of the
smooth barriers. Both the frequency and the amplitude of the
oscillations become higher as the strength of the MBs is increased (Fig. \ref%
{fig3:Conductance}). Furthermore, the frequency decreases as the length of the MB is
decreased (not shown). This behavior is similar to the conductance resonances in quantum
point contacts with abrupt openings \cite{Maao1994} and originates from
multiple reflections at the edges of the MB along the transport direction. The multiple
reflections at the edges lead to the $\mathrm{{Fabry-P\acute{e}rot}}$ type oscillations,
as can be seen in Fig. \ref{fig4:LDOSxE} for the case of a GNR with zigzag edges where
the number of maxima in the LDOS along the transport direction changes by one for
successive resonances. Similar to the case of a smooth quantum point contact
\cite{Maao1994}, a gradual change of the magnetic field reduces the reflection
probabilities and suppresses the resonances, resulting in smaller oscillation amplitudes.

\begin{figure*}[tbp]
\includegraphics[scale=1.0]{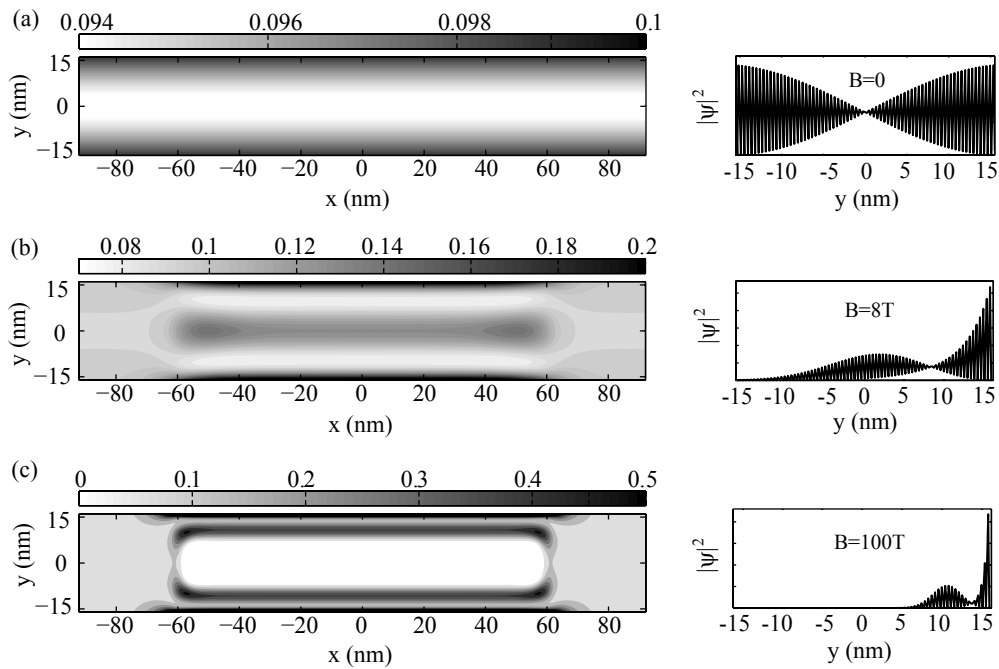}
\caption{Spatial distributions of local density of states in the zigzag GNR with $N=151$ (left column)
without a magnetic barrier (a), with a rectangular barrier of
strength $B=8\,\mathrm{T}$ (b) and $B=100\,\mathrm{T}$ (c) extending from $%
x=-60\,\mathrm{nm}$ to $x=60\,\mathrm{nm}$. The gray scales (indicated by the bars on top) give the LDOS in units of
$t^{-1}_{0}$nm$^{-2}$ .  The Fermi energy is $E_F=0.01t_0$ corresponding to the case when
only the lowest propagating state is occupied.The corresponding probability densities of the lowest Bloch state
 (in arbitrary units) in an infinite homogeneous wire are shown in the right column.} \label{fig5:LDOSxy}
\end{figure*}

The dependence of the conductance on the magnetic barrier suggests that complete
confinement by magnetic barriers is not possible due to the presence of the lowest
propagating state, in stark contrast to the case of two-dimensional graphene
sheets.\cite{Martino} To shed more light on the influence of the MB on the lowest
propagating state, we study the local density of state (LDOS) of the zigzag GNR in the
energy interval where only the lowest propagating state is occupied, see Fig.
\ref{fig5:LDOSxy}. A rectangular MB strongly modifies the lowest
propagating state in the transverse direction.  The
wave function patterns in the barrier region can be easily understood from analysis of the
corresponding patterns of Bloch states in the homogeneous wire. The latter are shown in
the right column of  Fig. \ref{fig5:LDOSxy}.  The lowest propagating state in the absence of the MB (Fig.
\ref{fig5:LDOSxy} (a)) extends across the whole GNR at this energy. At $B=8\,\mathrm{T}$,
its probability density has a node at about $7\,\mathrm{nm}$ away from the edge, and a
local maximum is formed close to the center of the GNR. As $B$ increases, this structure
is pushed towards the edge of the GNR while its shape persists. A comparison of these patterns
demonstrates that the wave function in the barrier region is directly related to the
corresponding eigenstate of the homogeneous channel. Note that due to reflection on the
barrier boundaries the edge state circulates inside the barrier region such that in this
region $|\psi|^2$ has similar amplitude near the upper and lower edges of the wire. We note further that the rapid
oscillations corresponding to the wave functions on the sites belonging to the A and B sublattices are averaged out in the grey-scale plots to the left.

The presence of this lowest propagating state apparently hampers the control
of the carrier confinement in GNRs by MBs. It would thus be important to
find a way to localize the lowest propagating state in gapless GNRs. There
has been a lot of theoretical effort to explore a way to open a bandgap in
metallic GNRs, such as application of uniaxial strain, boron doping, and
introduction of a line of impurities \cite{Filho}. To the best of our
knowledge, no metallic behavior of GNRs with widths as studied here has been
observed experimentally \cite{Han}. It was pointed out that the major
discrepancies between experiments and theory may arise from the assumptions
of perfect GNRs with a well-defined type of edge used in most theoretical
studies. Experimental observations reveal that edge disorder is very
significant on natural graphite edges and etched GNRs. Theoretical studies
have shown that the edge disorder dramatically affects the transport
properties and may turn the metallic ribbons into semiconductors \cite%
{Querlioz}. Since the edge disorder is usually present in realistic GNRs, we
investigate the transport properties in such GNRs subjected to a MB.

\begin{figure}[tbp]
\includegraphics[scale=1.0]{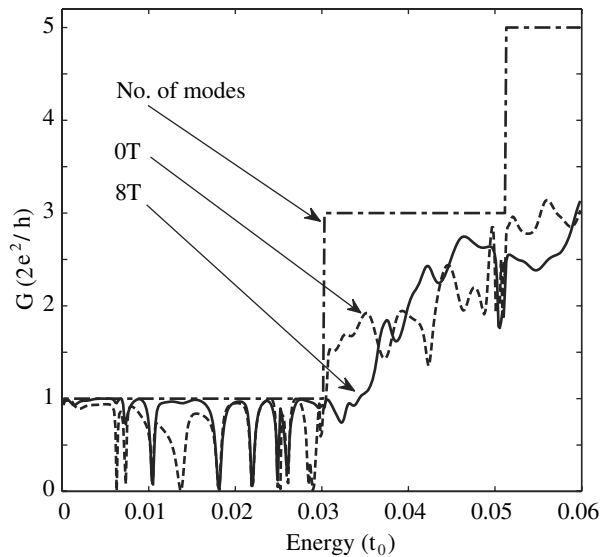}
\caption{The Fermi energy dependence of the conductance of defective
graphene ribbons with and without the rectangular MB of strengths $8\,%
\mathrm{T}$. The edge defect concentration is 30\%.}
\label{fig6:GNRdisorder}
\end{figure}

In Fig.  \ref{fig6:GNRdisorder} we show the conductance of a zigzag GNR with edge
disorder with and without a MB as a function of Fermi energy. The edge defects are
implemented by randomly removing 30 \% of the atoms at the edges on both sides of the
GNR, both inside and outside the magnetic barrier region. We first look at the
conductance behavior in low energy regime. The characteristic feature is the appearance
of conductance dips at the specific values of the Fermi energy. Similar dips are also
found in GNRs with additional bonds attached to the edges \cite{Wakabayashi}. As the
concentration of edge defects increases, the dips become more prominent and more
zero-conductance dips appear, and their position and structure changes as the defect
configuration is varied (not shown). When the MB is activated, the position of some
conductance dips, presumably those originating from defects underneath the MB, move in
energy while their amplitude is suppressed. The effect of the magnetic field is therefore
to delocalize the lowest propagating state which have been localized by the edge defects.
These results suggest that a magnetic barrier can in fact be used to switch the
conductance in a GNR, but the mechanism differs from that one to be expected for magnetic
barriers in two-dimensional graphene. Activation of the MB is able to delocalize the
lowest propagating state in GNRs with disorder, thereby switching the conductance from
zero to one.

\section{Summary and conclusions}

We have provided an extensive theoretical study of the transport through graphene
nanoribbons under the influence of magnetic barriers. The magnetic barrier modifies the
energy spectrum of the nanoribbon locally, which results in an energy shift of the
conductance steps. In addition, multiple reflections along the transport direction
between the entrance and the exit of the magnetic barrier generate
$\mathrm{{Fabry-P\acute{e}rot}}$ resonances, the magnitude of which depends on the
gradient of the magnetic field. These $\mathrm{{Fabry-P\acute{e}rot}}$ resonances are
strongly suppressed in the case of magnetic barriers with smooth confinement. The lowest
propagating state present in zigzag and metallic armchair GNRs is only weakly modified by
magnetic barriers of realistic strengths. However, localization of the lowest propagating
state by disorder can be lifted by a perpendicular magnetic field, which offers a concept
for magnetic barrier induced conductance switching in GNRs with disordered edges.

In this paper we also present a novel method based on the Greens function technique for
the calculation of the magnetosubband structure, Bloch states and magnetoconductance of
the graphene nanoribbons in a perpendicular magnetic field. The non-trivial part of the
method is the calculation of the surface Greens function $\Gamma $, which typically
requires very time-consuming self-consistent calculations. We, however, introduced a
novel way to calculate the surface Greens function that does not require self-consistent
calculations, \cite{EPAPS}and where $\Gamma $ is simply obtained from the solutions of the
eigenequation of the dimension $2N\times 2N$ (with $N$ being the width of the
nanoribbon). Utilization of this method obviously greatly facilitates computations,
making the present method by far more efficient in comparison to the existing methods
based on the self-consistent calculations of $\Gamma $. The programming codes are freely available in the AIP EPAPS electronic depository.\cite{EPAPS}

\begin{acknowledgments}
The authors would like to thank W. H\"ausler and R. Egger for fruitful
discussions. H.X. and T.H. acknowledge financial support from the
Heinrich-Heine Universit\"{a}t D\"{u}sseldorf and from the German Academic
Exchange Service (DAAD) within the DAAD-STINT collaborative grant.
\end{acknowledgments}

\appendix

\section{Hopping matrixes U}

In this appendix we provide explicit expressions for hopping matrixes $%
U_{i,i^{\prime }}$, Eq. (\ref{matrix elements}), for armchair and zigzag
ribbons in the Landau gauge $\mathbf{A}=(-By,0)$. The numbering of slices
and sites, $r=i,j,$ within a unit cell is given in Fig. (\ref{fig2:Bloch}),
and the definition of the phases $\theta _{r,r+\Delta }=2\pi \phi
_{r,r+\Delta }/\phi _{0}$ and the corresponding line integrals $\phi
_{r,r+\Delta }$ are given by Eq. (\ref{Phase}). In the expressions given
below $y_{i,j}$ stands for the $y$-coordinate of the site ($i,j$).

\begin{widetext}
\subsubsection{Armchair graphene ribbon}
\begin{equation}
\left( U_{1,0}\right) _{j,j^{\prime }}=-t_{0}\exp (i\theta _{0,j;1,j})\delta
_{j,j^{\prime }};\ \ U_{0,1}=U_{1,0}^{+}
\end{equation}%
where $\phi _{0,j;1,j}=-By_{0,j}a;$
\begin{equation}
U_{2,1}=-t_{0}%
\begin{pmatrix}
e^{i\theta _{1,1;2,1}} & e^{i\theta _{1,2;2,1}} &  &  &  \\
& e^{i\theta _{1,2;2,2}} & \ddots  & 0 &  \\
&  & \ddots  & e^{i\theta _{1,N-2;2,N-1}} &  \\
& 0 &  & e^{i\theta _{1,N-1;2,N-1}} & e^{i\theta _{1,N-1;2,N}} \\
&  &  &  & e^{i\theta _{1,N;2,N}}%
\end{pmatrix}%
;\quad U_{1,2}=U_{2,1}^{+}
\end{equation}%
where $\phi _{1,j;2,j}=-\frac{B}{2}\left( y_{1,j}a+\frac{\sqrt{3}}{4}%
a^{2}\right) ,$ $\phi _{1,j+1;2,j}=-\frac{B}{2}\left( y_{1,j+1}a-\frac{\sqrt{%
3}}{4}a^{2}\right) ;$
\begin{equation}
\left( U_{3,2}\right) _{j,j^{\prime }}=-t_{0}\exp (i\theta _{2,j;3,j})\delta
_{j,j^{\prime }};\ \ U_{2,3}=U_{3,2}^{+}
\end{equation}%
where $\phi _{2,j;3,j}=-By_{2,j}a;$%
\begin{equation}
U_{4,3}=-t_{0}%
\begin{pmatrix}
e^{i\theta _{3,1;4,1}} &  &  &  &  \\
e^{i\theta _{3,1;4,2}} & e^{i\theta _{3,2;4,2}} &  & 0 &  \\
& e^{i\theta _{3,2;4,3}} & \ddots  &  &  \\
& 0 & \ddots  & e^{i\theta _{3,N-1;4,N-1}} &  \\
&  &  & e^{i\theta _{3,N-1;4,N}} & e^{i\theta _{3,N;4,N}}%
\end{pmatrix}%
;\quad U_{3,4}=U_{4,3}^{+}
\end{equation}%
where $\phi _{3,j;4,j}=-\frac{B}{2}\left( y_{3,j}a-\frac{\sqrt{3}}{4}%
a^{2}\right) ,\;\phi _{3,j;4,j+1}=-\frac{B}{2}\left( y_{3,j}a+\frac{\sqrt{3}%
}{4}a^{2}\right) ;$ and, because of periodicity,
\begin{equation}
U_{5,4}=U_{1,0,}\;U_{4,5}=U_{0,1}.
\end{equation}
\subsubsection{Zigzag graphene ribbon}
\begin{equation}
\left( U_{1,0}\right) _{j,j^{\prime }}=-t_{0}\exp (i\theta _{0,j;1,j})\delta
_{j,j^{\prime }};\ \ U_{0,1}=U_{1,0}^{+},
\end{equation}%
where for odd $j$: $\phi _{0,j;1,j}=-\frac{\sqrt{3}}{2}B\left( y_{0,j}a+%
\frac{1}{4}a^{2}\right) ,$ and for even $j$: $\phi _{0,j;1,j}=-\frac{\sqrt{3}%
}{2}B\left( y_{0,j}a-\frac{1}{4}a^{2}\right) $%
\begin{equation}
\left( U_{2,1}\right) _{j,j^{\prime }}=-t_{0}\exp (i\theta _{1,j;2,j})\delta
_{j,j^{\prime }};\ \ U_{1,2}=U_{2,1}^{+},
\end{equation}%
where for odd $j$:  $\phi _{1,j;2,j}=-\frac{\sqrt{3}}{2}B\left( y_{1,j}a-%
\frac{1}{4}a^{2}\right) ,$ and for even $j$: $\phi _{1,j;2,j}=-\frac{\sqrt{3}%
}{2}B\left( y_{1,j}a+\frac{1}{4}a^{2}\right) ;$ and, because of periodicity,
\begin{equation}
U_{3,2}=U_{1,0,}\;U_{2,3}=U_{0,1}.
\end{equation}

\end{widetext}

\end{document}